\documentclass[aps,pra,showpacs,twocolumn,amsmath,superscriptaddress]{revtex4-1}
\usepackage{bm}
\usepackage{tikz}  
\usepackage{pgf}
\usepackage{graphicx,epsfig,subfigure,dsfont,amssymb,amsmath,amsthm,amsfonts,amsbsy,mathrsfs,amscd}
\usepackage{epstopdf}
\usepackage{amsmath}
\usepackage{bm}
\usepackage{latexsym}
\usepackage{amssymb} 
\usepackage{amsthm} 
\usepackage{times} 
\usepackage{graphicx}

\newcommand{\njp}{New. J. Phys.~}

\newcommand{\tinyspace}{\mspace{1mu}}

\newcommand{\abs}[1]{\left\lvert\tinyspace #1 \tinyspace\right\rvert}

\newcommand{\proj}[1]{| #1\rangle\!\langle #1 |}

\renewcommand{\t}{{\scriptscriptstyle\mathsf{T}}}

 \newcommand{\var}{\operatorname{Var}}

\def\1{\mathbf{1}}

\newcommand{\iinner}[2]{\langle #1 | #2\rangle}
\newcommand{\out}[2]{| #1\rangle\langle #2 |}

\newcommand{\Innerm}[3]{\left\langle #1 \left| #2 \right| #3 \right\rangle}

\newcommand{\Pa}[1]{\left(#1\right)}

\newcommand{\Br}[1]{\left[#1\right]}
\newcommand{\set}[1]{\{#1\}}
\newcommand{\Set}[1]{\left\{#1\right\}}

\newcommand{\ket}[1]{|#1\rangle}

\DeclareMathOperator{\trace}{Tr}

\newcommand{\Ptr}[2]{\trace_{#1}\Pa{#2}}

\newcommand{\Tr}[1]{\Ptr{}{#1}}

\def\bsA{\boldsymbol{A}}\def\bsB{\boldsymbol{B}}

\def\bsP{\boldsymbol{P}}

\def\bsr{\boldsymbol{r}}

\theoremstyle{definition}

\begin{document}


\title{Standard symmetrized variance with applications to  coherence, uncertainty and entanglement}


\author{Ming-Jing Zhao}
\affiliation{School of Science, Beijing Information Science and
Technology University, Beijing, 100192, PR~China}
\affiliation{State key Laboratory of Networking and Switching Technology (Beijing University of Posts and Telecommunications), Beijing, 100876, PR~China}

\author{Lin Zhang}
\email{E-mail: godyalin@163.com}
\affiliation{Institute of Mathematics, Hangzhou Dianzi University,
Hangzhou 310018, China}

\author{Shao-Ming Fei}
\affiliation{School of Mathematical Sciences, Capital Normal
University, Beijing 100048, China}
\affiliation{Max-Planck-Institute
for Mathematics in the Sciences, 04103 Leipzig, Germany}

\begin{abstract}
Variance is a ubiquitous quantity in quantum information theory.
Given a basis, we consider the averaged variances of a fixed
diagonal observable in a pure state under all possible permutations
on the components of the pure state and call it the symmetrized
variance. Moreover we work out the analytical
expression of the symmetrized variance and find that such expression
is in the factorized form where two factors separately depends on the
diagonal observable and quantum state.
By shifting the factor corresponding to the diagonal observable, we introduce the notion
named the standard symmetrized variance for the pure state which is
independent of the diagonal observable. We then extend the standard
symmetrized variance to mixed states in three different ways, which
characterize the uncertainty, the coherence and the coherence of
assistance, respectively. These quantities are evaluated
analytically and the relations among them are established. In
addition, we show that the standard symmetrized variance is also an
entanglement measure for bipartite systems. In this way, these
different quantumness of quantum states are unified by the variance.
\end{abstract}

\maketitle

\section{Introduction}

Measuring statistically the deviation of
measurement outcomes from the ideal value of quantum measurements on
given quantum states, the variance plays an important role in quantum
physics and quantum information theory. The first uncertainty
relation known as the Heisenberg's uncertainty relation was given
in terms of the variance \cite{W. Heisenberg,H. P. Robertson}, which
describes the restrictions on the accuracy of measurement results of two
or more noncommutative observables. Since then, many efforts are
made to find tighter and state-independent lower bounds on such kind
of uncertainty relations \cite{Pati2014,S. Friedland,Schwonnek2017}.
In order to study the restriction between the uncertainties for two
or more observables, the notion of the uncertainty region is also
put forward and characterized \cite{Werner2015,Zhang2021}.

For any given mixed state and observable, it has been shown that the
minimal averaged variance among all possible pure-state
decompositions is exactly the quantum Fisher information and the
maximal averaged variance is the variance itself
\cite{Yu2013,Toth2013,S. Braunstein,G. Toth2017}. Concerning the
relation between the variance and quantum Fisher information, the
improved uncertainty relations have been derived in terms of quantum
Fisher information and variance \cite{S. Luo2005pra,Toth2021,Gessner2021}.
Moreover, the variance-based uncertainty relation has been
incorporated into quantum multiparameter estimation by investigating
the quantum Fisher information matrix and the classical Fisher
information matrix from measurements \cite{Lu2021}.
It is well known
that the quantum Fisher information places the fundamental limit to
the accuracy for estimating an unknown parameter and plays an important role in quantum metrology.
The uncertainty
relation and quantum metrology are then closely connected by the
variance and quantum Fisher information.

More than that, the variance can also be used to detect quantum
entanglement \cite{Schwonnek2017,H. F. Hofmann,O. Guhne,N. Li2013,Y.
Hong} and quantify quantum coherence \cite{Y.
Zhang2020,Li2021,Xu2021}. In view of the importance of the variance
in quantum information theory, we aim to explore the role played by
the variance,  {as a unifying tool}, in characterizing
the quantumness. In a $d$-dimensional system with fixed
basis $\set{\ket{i}:i=0,1,\ldots,d-1}$, any pure state can be
expressed as $\ket{\psi}=\sum_{i=0}^{d-1} \psi_i \ket{i}$. Since the
quantumness of the pure state is invariant under relabeling the
coefficients $\set{\psi_i}$, in order to study the quantumness of
the pure state $\ket{\psi}$, one may consider the averaged
quantumness in the set of pure states $\set{\bsP_{\pi}\ket{\psi}}$,
where $\bsP_{\pi}$ are all possible permutation matrices
corresponding to the permutations $\pi$ of the permutation group
$S_d$. In this context, we propose the averaged variances of any
diagonal observable over the set of $\set{\bsP_{\pi}\ket{\psi}}$ and
named as symmetrized variance $\widehat{\var}(\ket{\psi}, \bsA)$.
By analysis and calculation, we obtain the analytical expression of the symmetrized variance and find that the information of the pure state and the diagonal observable are factorized.
By shifting the information of the diagonal observable, we introduce the core notion of this paper called {\it the standard symmetrized variance} $\widehat{V}(|\psi\rangle)$.

\begin{figure}[ht]\centering
{\begin{minipage}[b]{1\linewidth}
\includegraphics[width=1\textwidth]{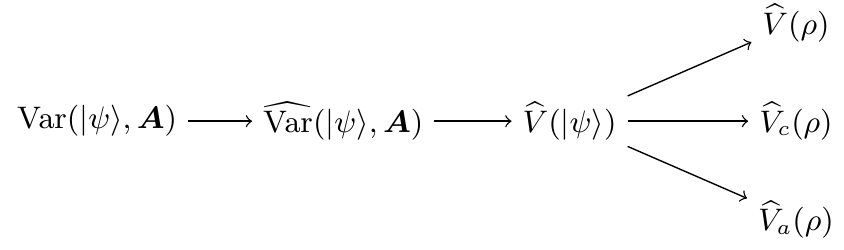}
\end{minipage}}





\caption{The diagram of the standard symmetrized variances.
$\var(\ket{\psi}, \bsA)$ is the variance of the diagonal observable
$\bsA$ in the pure state $\ket{\psi}$ defined in Eq. \eqref{eq var
pure}. $\widehat{\var}(\ket{\psi}, \bsA)$ is the symmetrized
variance of the diagonal observable $\bsA$ in the pure state
$\ket{\psi}$ defined in Eq. \eqref{eq varhat pure}.
$\widehat{V}(\ket{\psi})$ is the standard symmetrized variance of
the pure state $\ket{\psi}$ defined in Eq. \eqref{eq vpsi}.
$\widehat{V}(\rho)$ is the standard symmetrized variance of the
mixed state $\rho$ defined in Eq. \eqref{eq vrho}, which is an
uncertainty measure. $\widehat{V}_c(\rho)$ is the convex roof
extension of the standard symmetrized variance defined in Eq.
\eqref{eq vcrho}, which is a coherence measure.
$\widehat{V}_a(\rho)$ is the concave bottom extension of the
standard symmetrized variance defined in Eq. \eqref{eq varho}, which
is the coherence of assistance. }\label{sch}
\end{figure}

Although the standard symmetrized variance is  {in
terms of variance}, we  {find} that it is independent
of the diagonal observable and just  {depends on the
pure state itself}. Namely, the standard symmetrized variance describes the quantumness in the pure state. We generalize the standard
symmetrized variance from the pure states to mixed states based on
three different extensions. The first extension is the standard
symmetrized variance $\widehat{V}(\rho)$ obtained by replacing the pure state
with mixed ones directly. We get the analytical formula of
$\widehat{V}(\rho)$ and show that it is a measure of uncertainty
associated to the state $\rho$. The second extension is the convex roof
extension $\widehat{V}_c(\rho)$. The lower bound of
$\widehat{V}_c(\rho)$ is derived in terms of the eigenvalues and
eigenvectors of the mixed state, which can be attained for qubit
quantum states. The convex roof extension $\widehat{V}_c(\rho)$ is a
coherence measure for $\rho$. The third extension is the concave
bottom extension $\widehat{V}_a(\rho)$, which, dual to
$\widehat{V}_c(\rho)$, is the coherence of assistance. These three
extensions are depicted in Bloch ball for any qubit state. The
standard symmetrized variance can also give rise to an entanglement
measure in bipartite systems. The schematic on the relations among
these quantities is given in Fig. \ref{sch}.

\section{The standard symmetrized variance}

For any quantum state $\rho$ and an observable $\bsA$ in a $d$-dimensional system, the variance
\begin{eqnarray*}
\var(\rho, \bsA)=\Tr{\bsA^2\rho} - [\Tr{\bsA\rho}]^2
\end{eqnarray*}
is a measure of uncertainty of the observable $\bsA$ in the state
$\rho$.
In this paper, we fix the reference basis as $\set{\ket{i}:i=0,1,\ldots,d-1}$
and just consider the variance of diagonal
observables $\bsA$ in quantum state $\rho$ throughout this paper.

\subsection{Standard symmetrized variance for pure states}

For any pure state $\ket{\psi}=\sum_{i=0}^{d-1}  \psi_i \ket{i}$,
the vector $\Pa{r_0, r_1, \ldots, r_{d-1}}^\t$ with
$r_i:=\abs{\psi_i}^2$ is called the \emph{coherence vector} of a
pure state $\ket{\psi}$ \cite{S. Du,H. Zhu}. Based on the variance
\begin{eqnarray}\label{eq var pure}
\var(\ket{\psi}, \bsA)&=&\Innerm{\psi}{\bsA^2}{\psi} -
\Innerm{\psi}{\bsA}{\psi}^2,
\end{eqnarray}
we define {\it the symmetrized variance} of the observable $\bsA$ in the
pure state $\ket{\psi}$ as
\begin{eqnarray}\label{eq varhat pure}
\widehat{\var}(\ket{\psi}, \bsA) &=&
\frac1{d!}\sum_{\pi\in S_d} \var( \bsP_{\pi}\ket{\psi},  \bsA)
\\&=&\frac1{d!}\sum_{\pi\in S_d}
\var(\ket{\psi},   \bsP^\dagger_{\pi} \bsA \bsP_{\pi})\nonumber.
\end{eqnarray}
$\widehat{\var}(\ket{\psi}, \bsA)$ is the averaged variances of the
observable $\bsA$ in the set of pure states $\set{\bsP_{\pi}
\ket{\psi}}$.

The symmetrized variance $\widehat{\var}(\ket{\psi}, \bsA)$
satisfies the following properties: (i) $\widehat{\var}(\ket{\psi},
\bsA)$ is symmetric under relabeling the entries of the vector
$\{\psi_i\}$; (ii) $\widehat{\var}(\ket{\psi},
\bsA)$ is a concave function of the state $\ket{\psi}$; (iii)
$\widehat{\var}(\ket{\psi}, \bsA)=\widehat{\var}(\ket{\psi},
\bsP^\dagger_{\pi} \bsA \bsP_{\pi})$ for any permutation matrix
$\bsP_{\pi}$.

{\bf Theorem 1.}\label{prop pure analy}
For any pure state $\ket{\psi}=\sum_{i=0}^{d-1}  \psi_i \ket{i}$ and
diagonal observable $\bsA$, the symmetrized variance
$\widehat{\var}(\ket{\psi}, \bsA)$ is given by
\begin{eqnarray}
\widehat{\var}(\ket{\psi}, \bsA) =\kappa(\bsA)
\Br{1-\sum^{d-1}_{i=0}|\psi_i|^4}.
\end{eqnarray}
with $\kappa(\bsA)=\frac{d\Tr{\bsA^2}-[\Tr{\bsA}]^2}{d(d-1)}$.

The proof is in the Appendix A.
Theorem 1 provides an analytical expression of
the symmetrized variance. This formula shows that the information
about the diagonal observable $\bsA$ and the pure
state $|\psi\rangle$ given in the symmetrized variance
$\widehat{\var}(\ket{\psi}, \bsA)$ are factorized. Inspired by this,
we define {\it the standard symmetrized variance} of the pure state
$\ket{\psi}$ as
\begin{eqnarray}\label{eq vpsi}
\widehat{V}(\ket{\psi})=\frac{1}{\kappa(\bsA)}\widehat{\var}(\ket{\psi},
\bsA),
\end{eqnarray}
which can be expressed as
\begin{equation}
\widehat{V}(\ket{\psi})=1-\sum^{d-1}_{i=0}\abs{\psi_i}^4.
\end{equation}
This standard symmetrized variance $\widehat{V}(\ket{\psi})$ is
independent of the observable $\bsA$ and is only given by the pure
state $\ket{\psi}$. The standard symmetrized variance
$\widehat{V}(\ket{\psi})$ has the following properties. First,
$\widehat{V}(\ket{\psi})$ is symmetric and concave with respect to
the coherence vector $(r_0, r_1,\cdots, r_{d-1})^\t$. Second,
$0\leqslant \widehat{V}(\ket{\psi})\leqslant 1-\frac{1}{d}$.
Moreover, $\widehat{V}(\ket{\psi})=0$ if and only if $\ket{\psi}$ is
incoherent, namely, $\ket{\psi}\langle\psi|$ is diagonal under the reference basis.
$\widehat{V}(\ket{\psi})= 1-\frac{1}{d}$ if and only if $\ket{\psi}$
is maximally coherent, $\ket{\psi}=\frac{1}{\sqrt{d}}
\sum_{j=0}^{d-1} e^{i\theta_j}\ket{j}$.

\subsection{The extension to mixed states}

Now we extend the standard symmetrized variance
$\widehat{V}(\ket{\psi})$ from pure states to mixed states in three
different ways.

(1) {\it The standard symmetrized variance $\widehat{V}(\rho)$ and
uncertainty.} For mixed states, we define the standard symmetrized
variance $\widehat{V}(\rho)$ as
\begin{eqnarray}\label{eq vrho}
\widehat{V}(\rho)=\frac{1}{\kappa(\bsA)}\frac1{d!}\sum_{\pi\in S_d}
\var( \bsP_{\pi} \rho \bsP^\dagger_{\pi},  \bsA ).
\end{eqnarray}
Let $\Pi=\Set{\Pi_i=\out{i}{i}: i=0,1,\cdots, d-1}$ be the
projective measurements, $\Pi(\rho)=\sum_i\Pi_i \rho \Pi_i$ be the
post-measurement state, and $S_L(\rho)=1-\Tr{\rho^2}$ be the linear
entropy which measures the mixedness of $\rho$. Then the standard
symmetrized variance $\widehat{V}(\rho)$ actually gives the
mixedness of postmeasurement state $\Pi(\rho)$.

{\bf Theorem 2.} For any quantum state $\rho$, the standard symmetrized variance
$\widehat{V}(\rho)$ is given by
\begin{eqnarray}
\widehat{V}(\rho)= S_L(\Pi(\rho)).
\end{eqnarray}

The proof is in the Appendix B.
The standard symmetrized variance $\widehat{V}(\rho)$ has the
following properties. (i) It is a concave function of $\rho$; (ii)
It is invariant under the permutation $\bsP_{\pi}$,
$\widehat{V}(\rho)=\widehat{V}(\bsP_{\pi} \rho \bsP^\dagger_{\pi})$;
(iii) $0\leqslant \widehat{V}(\rho)\leqslant 1-\frac{1}{d}$.
$\widehat{V}(\rho)=0$ if and only if $\rho$ is pure and incoherent.
$\widehat{V}(\rho)= 1-\frac{1}{d}$ if and only if $\rho$ is
maximally coherent; (iv) For $\rho=p_1 \rho_1 \oplus p_2 \rho_2
\oplus \cdots \oplus p_n \rho_n$, it satisfies
$\widehat{V}(p_1 \rho_1 \oplus p_2 \rho_2\oplus \cdots \oplus p_n
\rho_n)= p_1^2\widehat{V}(\rho_1) + p_2^2\widehat{V}(\rho_2)+ \cdots
+ p_n^2\widehat{V}(\rho_n) + \sum_{i\neq j}p_ip_j$, which is proved in the Appendix C.

Since the standard symmetrized variance $\widehat{V}(\rho)$ is a
concave function of the vector $(\rho_{00},\ldots,\rho_{d-1,d-1})^\t$,
where $\rho_{ii}=\Innerm{i}{\rho}{i}$, and is invariant under the
permutation $\bsP_{\pi}$, $\widehat{V}(\rho)$ is a measure of
uncertainty which quantifies the uncertainty associated with the
quantum state $\rho$ in the framework of uncertainty measure
\cite{S. Friedland}. In Refs. \cite{S. Luo2005,S. Luo2017,Y.
Sun2021} the authors proposed the idea to split the uncertainty into
quantum part and classical part. In Refs. \cite{S. Luo2005,S.
Luo2017} the total uncertainty is specified to the variance of
observable $\bsA$ in state $\rho$, while the quantum uncertainty is
specified to the Wigner-Yanase skew information. Here we can
decompose the standard symmetrized variance $\widehat{V}(\rho)$ into
the quantum uncertainty $Q(\rho)$ and the classical uncertainty
$C(\rho)$,
\begin{eqnarray}\label{uncertainty decomp}
\widehat{V}(\rho)=C(\rho) + Q(\rho),
\end{eqnarray}
where $C(\rho)=S_L(\rho)$ is the mixedness of $\rho$,
$Q(\rho)=S_L(\Pi(\rho))-S_L(\rho)$ is just the genuine coherence of
$\rho$ \cite{Y. Sun2021,J. Vicente}.

Theorem 2 can also be viewed as an uncertainty
relation in summation form if we reexpress the standard symmetrized
variance $\widehat{V}(\rho)$ in the form of variance. Since
$\widehat{V}(\rho)$ can be equivalently rewritten as $\sum_{\pi\in
S_d} \var( \rho , \bsP^\dagger_{\pi} \bsA
\bsP_{\pi})=d!\kappa(\bsA)S_L(\Pi(\rho))$, it can be viewed as an
uncertainty relation among the observables $\set{\bsP^\dagger_{\pi}
\bsA \bsP_{\pi}}_{\pi}$ in the state $\rho$. For example, for qubit
systems this uncertainty relation reduces to $\var( \rho , \bsA )+
\var( \rho , \sigma_1 \bsA \sigma_1)=[2
\Tr{\bsA^2}-[\Tr{\bsA}]^2] \widehat{V}(\rho)$, which gives an
uncertainty relation for the commutative diagonal observables $\bsA$ and $\sigma_1 \bsA \sigma_1$ in the form of equality, with $\sigma_1=|0\rangle\langle1|+|1\rangle\langle0|$.

(2) {\it The convex roof extension $\widehat{V}_c(\rho)$ and
coherence measure.} Now we extend the standard symmetrized variance
to mixed states by convex roof extension,
\begin{eqnarray}\label{eq vcrho}
\widehat{V}_c(\rho)&=&\min_{\set{p_k, \ket{\psi_k}}} \sum_k p_k
\widehat{V}(\ket{\psi_k}),
\end{eqnarray}
where the minimum runs over all possible pure state ensemble
decompositions of $\rho=\sum_k p_k \out{\psi_k}{\psi_k}$.

Since the standard symmetrized variance $\widehat{V}(\ket{\psi})$ is
a real symmetric concave function of the coherence vector $\Pa{r_0,
r_1, \ldots, r_{d-1}}^\t$, and is zero if and only if $\ket{\psi}$
is incoherent, $\widehat{V}_c(\rho)$ is a coherence measure \cite{S.
Du, H. Zhu}. The convex roof extension $\widehat{V}_c(\rho)$
satisfies the following properties. (i) $\widehat{V}_c(\rho)$ is
convex; (ii) $0\leqslant \widehat{V}_c(\rho)\leqslant
1-\frac{1}{d}$. Moreover, $\widehat{V}_c(\rho)=0$ if and only if
$\rho$ is incoherent, and $\widehat{V}_c(\rho)=1-\frac{1}{d}$ if and
only if $\rho$ is maximally coherent; (iii) For $\rho=p_1 \rho_1
\oplus p_2 \rho_2$, one has $\widehat{V}_c(p_1 \rho_1 \oplus p_2
\rho_2)={p_1}\widehat{V}_c(\rho_1) + {p_2} \widehat{V}_c(\rho_2)$ \cite{X. Yu}.

Now we evaluate the convex roof extension $\widehat{V}_c(\rho)$
analytically. Here we employ the following equality about the
minimum average invariance over all pure state decomposition and the
quantum Fisher information $ F(\rho, \bsA)= \sum_{k,l} 2
\frac{(\lambda_k-\lambda_l)^2}{\lambda_k+\lambda_l}\abs{A_{kl}}^2$,
where $A_{kl}=\Innerm{\phi_k}{\bsA}{\phi_l}$ with $\set{\lambda_k,
\ket{\phi_k}}$ the eigenvalues and the corresponding eigenstates of
the mixed state $\rho$ \cite{Yu2013,Toth2013,S. Braunstein,G.
Toth2017}, i.e.,
\begin{equation}\label{eq fisher and var}
F(\rho, \bsA)=4\min_{\set{p_k, \ket{\psi_k}}} \sum_k p_k
\var(\ket{\psi_k}, \bsA) .
\end{equation}

{\bf Theorem 3.}
For any mixed state $\rho$, the convex roof extension
$\widehat{V}_c(\rho)$ is bounded from below by
\begin{eqnarray}
\widehat{V}_c(\rho) \geqslant \frac{1}{2} \sum_{k,l}
\frac{(\lambda_k-\lambda_l)^2}{\lambda_k+\lambda_l} \sum_{i=0}^{d-1}
\abs{\phi^{(k)}_i}^2 \abs{\phi^{(l)}_i}^2,
\end{eqnarray}
where $\lambda_k$ and $\ket{\phi_k}=\sum_i \phi^{(k)}_i \ket{i}$ are
the eigenvalues and the corresponding eigenstates of the mixed state
$\rho$, respectively.

In particular, for qubit systems we have the following conclusion.

{\bf Theorem 4.}
For any qubit state $\rho$, the convex roof extension
$\widehat{V}_c(\rho)=\frac{1}{8} F(\rho, \sigma_3)
=\frac{1}{2}(\lambda_1-\lambda_2)^2
\abs{\Innerm{\phi_1}{\sigma_3}{\phi_2}}^2$, where $\set{\lambda_k,
\ket{\phi_k}}_{k=1,2}$ are the eigenvalues and the corresponding
eigenstates of $\rho$, $\sigma_3=\out{0}{0}-\out{1}{1}$ is the Pauli
matrix.

The proofs of Theorems 3 and 4 are in the Appendices D and E respectively.
By Theorem 4 we can see that the quantum Fisher information
is a coherence measure in qubit systems. But this is not true for
high dimensional systems, as a counterexample has been given in Refs.
\cite{X. Feng,H. Kwon}.

(3) {\it The concave bottom extension $\widehat{V}_a(\rho)$ and
coherence of assistance.} Now we extend the standard symmetrized
variance to mixed states by the concave bottom extension,
\begin{eqnarray}\label{eq varho}
\widehat{V}_a(\rho)&=&\max_{\set{p_k, \ket{\psi_k}}} \sum_k p_k
\widehat{V}(\ket{\psi_k}),
\end{eqnarray}
where the maximum runs over all possible pure state decompositions
of $\rho$.

The concave bottom extension $\widehat{V}_a(\rho)$ is dual to the
convex roof extension $\widehat{V}_c(\rho)$. Since the convex roof
extension  $\widehat{V}_c(\rho)$ is a coherence measure, the concave bottom extension $\widehat{V}_a(\rho)$ can be interpreted
operationally in the following way. Suppose Alice holds a state
$\rho^A$ with coherence $C(\rho^A)$. Bob holds another part of the
purified state of $\rho^A$. The joint state between Alice and Bob is
$\sum_k p_k \ket{\psi_k}_A \otimes \ket{k}_B$ with $\rho^A=\sum_k
p_k \proj{\psi_k}$. Bob performs local measurements $\set{\ket{k}}$
and informs Alice the measurement outcomes by classical
communication. Alice's quantum state will be in a pure state
ensemble $\set{ p_k,\ket{\psi_k}}$ with average coherence $\sum_k
p_k C(\proj{\psi_k})$. This process enables Alice to increase the
coherence from $C(\rho^A)$ to the average coherence $\sum_k p_k
C(\proj{\psi_k})$ due to the convexity of the coherence measure, and it is called the
assisted coherence distillation. The maximum average coherence is
called the coherence of assistance and
quantifies the one way coherence distillation rate \cite{E. Chitambar}. In this context,
the concave bottom extension $\widehat{V}_a(\rho)$ is a coherence of
assistance corresponding to the coherence measure standard
symmetrized variance.

The above three extensions of the standard symmetrized variance
satisfy the following relation,
\begin{eqnarray}
\widehat{V}_c(\rho)\leqslant\widehat{V}_a(\rho)\leqslant\widehat{V}(\rho).
\end{eqnarray}
The right inequality becomes equality if and only if there is a pure
state decomposition $\set{p_k,\ket{\psi_k}}$ of $\rho$ such that
$\Pi(\proj{\psi_k})=\Pi(\rho)$ for all $k$ \cite{M. Zhao}.
Especially, the equality $\widehat{V}_a(\rho) = \widehat{V}(\rho)$
holds true for all mixed states in two and three dimensional systems
according to the results in Ref. \cite{M. Zhao}.

For qubit systems, we have the following relation satisfied by
$\widehat{V}_a(\rho)$ and $\widehat{V}_c(\rho)$.

{\bf Theorem 5.}
For any qubit state $\rho$,
$$\widehat{V}_a(\rho)
-\widehat{V}_c(\rho)=\frac{3}{8}S_L(\Pi(\rho))+\frac{1}{4}S_L(\rho).$$

The proof is in the Appendix F.
Theorem 5 shows the convex roof extension
$\widehat{V}_c(\rho)$ coincides with the concave bottom extension
$\widehat{V}_a(\rho)$ if and only if $\rho$ is an incoherent pure
state. Hence, for all qubit mixed states, the concave bottom extension
$\widehat{V}_a(\rho)$ is strictly greater than the convex roof
extension $\widehat{V}_c(\rho)$. This not only gives an affirmative
answer to the conjecture that the concave bottom extension is strictly
greater than the convex roof extension for all mixed states and all coherence measures \cite{M.
Zhao}, but also implies that the coherence of all mixed states in
qubit systems can be increased in the assisted coherence
distillation \cite{M. Zhao}.

It is worthy to point out that the standard symmetrized variance
$\widehat{V}(\rho)$ as well as the convex roof extension
$\widehat{V}_c(\rho)$ and the concave bottom extension
$\widehat{V}_a(\rho)$ are all observable-independent. Therefore, in
order to evaluate these quantities theoretically or experimentally,
one may choose arbitrarily an appropriate diagonal observable. For
example, if we choose the diagonal observable $\bsA=\Pi_i$,
$0\leqslant i\leqslant d-1$, then the standard symmetrized variance
$\widehat{V}(\rho)=\frac{1}{d}\sum_{i=0}^{d-1} \var( \rho ,  \Pi_i)$
reduces to the uncertainty of $\rho$ proposed in Ref. \cite{Y.
Sun2021}, while the convex roof extension
$\widehat{V}_c(\rho)=\frac{1}{d} \min_{\set{p_k, \ket{\psi_k}}}
\sum_k p_k \sum_{i=0}^{d-1} \var( |\psi_k\rangle ,  \Pi_i )$ reduces
to the coherence measure proposed in Ref. \cite{Li2021}, and the
concave bottom extension is given by $\widehat{V}_c(\rho)=\frac{1}{d}
\max_{\set{p_k, \ket{\psi_k}}} \sum_k p_k \sum_{i=0}^{d-1} \var(
|\psi_k\rangle ,  \Pi_i )$ correspondingly.

\subsection{The standard symmetrized variance in qubit systems}

For qubit states, the density matrix can be expressed as
$\rho=\frac{1}{2}(I+\bsr\cdot \boldsymbol{\sigma})$, where
$\bsr=r(\sin\theta\cos\phi, \sin\theta\sin\phi,\cos\theta)$ is the
Bloch vector such that $r=\abs{\bsr}\in[0, 1]$ on Bloch ball,
$\boldsymbol{\sigma}=(\sigma_1,\sigma_2,\sigma_3)$ with the Pauli
matrices
$$
\sigma_1=\left(\begin{array}{cccccccc}
0 & 1 \\
1 & 0
\end{array}\right),\quad \sigma_2=\left(\begin{array}{cccccccc}
0 & -\mathrm{i} \\
\mathrm{i} & 0
\end{array}\right),\quad\sigma_3=\left(\begin{array}{cccccccc}
1 & 0 \\
0 & -1
\end{array}\right).
$$
Without loss of generality, we assume $\theta, \phi\in[0,\pi/2]$,
which means that the Bloch vector $\bsr$ is in the first octant of
the Bloch ball. By direct calculation, we have the convex roof
extension of the standard symmetrized variance
$\widehat{V}_{c}(\rho)$,
$$
\widehat{V}_{c}(\rho)=\frac{1}{2}r^2\sin^2\theta.
$$
The corresponding optimal pure state decomposition of $\rho$ which
gives rise to $\widehat{V}_{c}(\rho)$ is
$\Set{p_k^{(m)},\ket{\chi_k^{(m)}}}_{k=1}^2$, where
$$
\ket{\chi_1^{(m)}}=\cos\frac{\theta_m}{2}\ket{0}+e^{i\phi}\sin\frac{\theta_m}{2}\ket{1}
$$
with probability
$$
p_1^{(m)}=\frac{-1 + r^2 \sin^2{\theta} - r\cos{\theta} \sqrt{1 -
r^2 \sin^2{\theta}}}{-2 + 2 r^2 \sin^2{\theta}},
$$
and
$$
\ket{\chi_2^{(m)}}=\sin\frac{\theta_m}{2}\ket{0}+e^{\mathrm{i}\phi}
\cos\frac{\theta_m}{2}\ket{1}
$$
with probability
$$
p_2^{(m)}=\frac{-1 + r^2 \sin^2{\theta} + r\cos{\theta} \sqrt{1 -
r^2 \sin^2{\theta}}}{-2 + 2 r^2 \sin^2{\theta}},
$$
whose
Bloch vectors are $\bsr(\ket{\chi_1^{(m)}})=(\sin\theta_m\cos\phi,
\sin\theta_m\sin\phi,\cos\theta_m)$,
$\bsr(\ket{\chi_2^{(m)}})=(\sin(\pi-\theta_m)\cos\phi,
\sin(\pi-\theta_m)\sin\phi,\cos(\pi-\theta_m))$ with
$\theta_m=\arccos{\sqrt{1 - r^2 \sin^2{\theta}}}$.

\begin{figure}[ht]\centering
{\begin{minipage}[b]{1\linewidth}
\includegraphics[width=1\textwidth]{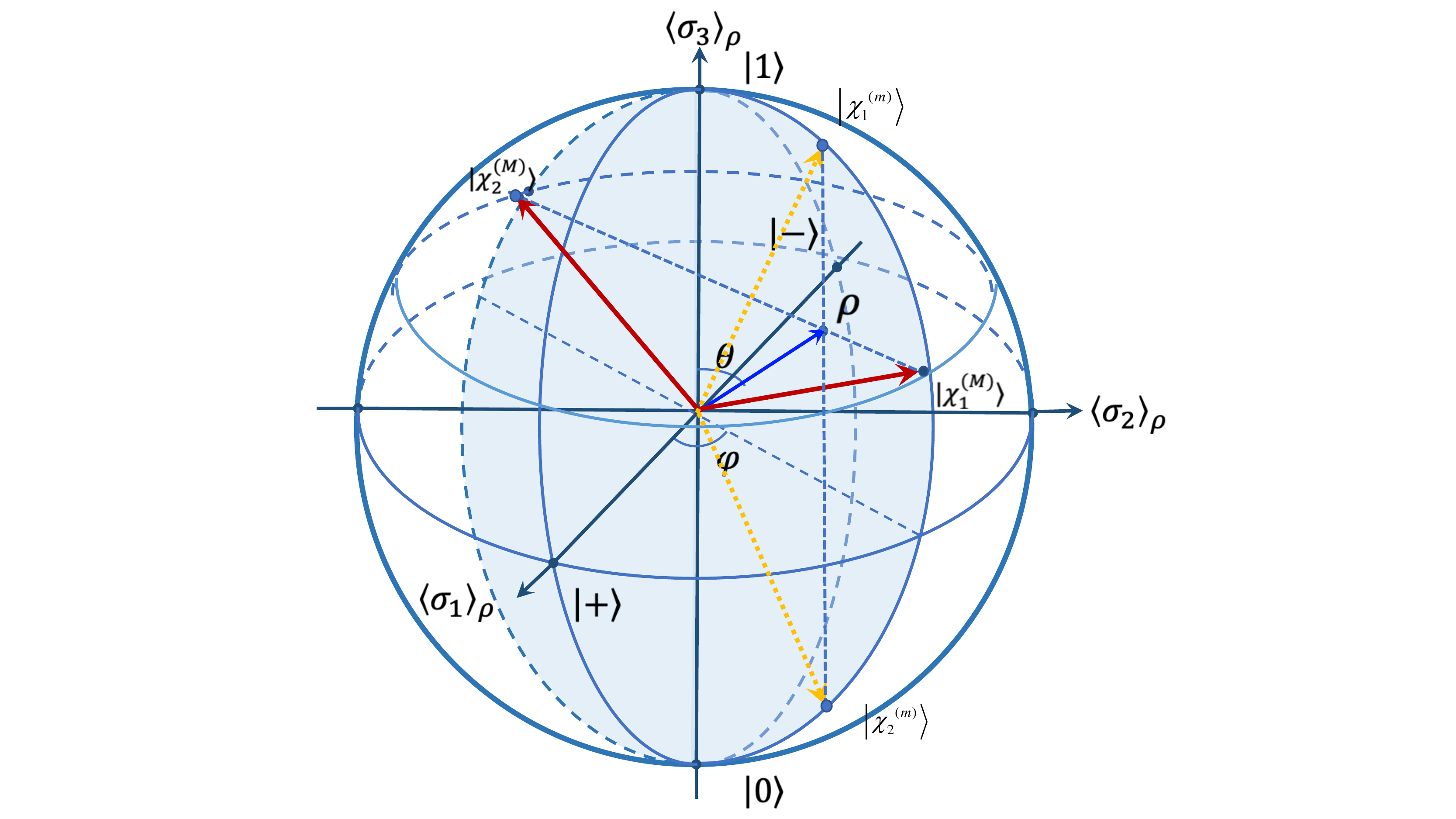}
\end{minipage}}
\caption{(Color Online) The optimal pure state decompositions of
$\rho$ for the convex roof extension $\widehat{V}_{c}(\rho)$ and the
concave bottom extension $\widehat{V}_{a}(\rho)$. The pure state
decomposition $\Set{p_k^{(m)},\ket{\chi_k^{(m)}}}_{k=1}^2$ (dashed
lines) is optimal for the convex roof extension
$\widehat{V}_{c}(\rho)$. The pure state decomposition
$\Set{p_k^{(M)}, \ket{\chi_k^{(M)}}}_{k=1}^2$ (solid lines) is
optimal for the concave bottom extension
$\widehat{V}_{a}(\rho)$.}\label{fig}
\end{figure}

The standard symmetrized variance $\widehat{V}(\rho)$ and the
concave bottom extension of the standard symmetrized variance
$\widehat{V}_{a}(\rho)$ are given by
$$
\widehat{V}(\rho)=\widehat{V}_{a}(\rho)=\frac{1}{2}+\frac{1}{2}r^2\cos^2\theta.
$$
The corresponding optimal pure state decomposition of $\rho$ which
attains the value of $\widehat{V}_{a}(\rho)$ is $\Set{p_k^{(M)},
\ket{\chi_k^{(M)}}}_{k=1}^2$, where
$$\ket{\chi_1^{(M)}}=\cos\frac{\theta_M}{2}\ket{0}
+e^{\mathrm{i}\phi}\sin\frac{\theta_M}{2}\ket{1}$$ with probability
$$
p_1^{(M)}=\frac{1}{2}\Pa{1+\frac{r\sin\theta}{\sin\theta_M}},
$$
and
$$
\ket{\chi_2^{(M)}}=\cos\frac{\theta_M}{2}\ket{0}
-e^{\mathrm{i}\phi}\sin\frac{\theta_M}{2}\ket{1}
$$
with probability
$$
p_2^{(M)}=\frac{1}{2}\Pa{1-\frac{r\sin\theta}{\sin\theta_M}},
$$
of
which the Bloch vectors are
$\bsr(\ket{\chi_1^{(M)}})=(\sin\theta_M\cos\phi,
\sin\theta_M\sin\phi,\cos\theta_M)$,
$\bsr(\ket{\chi_2^{(M)}})=(\sin\theta_M\cos(\pi+\phi),
\sin\theta_M\sin(\pi+\phi),\cos\theta_M)$ with $\theta_M=\arccos
(r\cos\theta)$, see Fig.~\ref{fig}.

\section{Connections with quantum entanglement}

To have an intuitive illustration on the relations between the
standard symmetrized variance and the quantum entanglement, let us consider bipartite
pure states in Schmidt decomposition form,
$\ket{\psi_{AB}}=\sum_{i=0}^{d-1} \lambda_i \ket{ii}$ with $\lambda_i\geq 0$, $\sum_i \lambda_i=1$ and local
diagonal observable $\bsA\otimes \bsB=\sum_i a_i \proj{i}\otimes
\sum_i b_i \proj{i}$ acting on $C^d \otimes C^d$. The
variance of the local diagonal observable $\bsA\otimes \bsB$ in bipartite pure state
$\ket{\psi_{AB}}$ is $\var(\ket{\psi_{AB}}, \bsA\otimes
\bsB)=\Innerm{\psi_{AB}}{\bsA^2 \otimes \bsB^2}{\psi_{AB}} -
\Innerm{\psi_{AB}}{\bsA\otimes \bsB}{\psi_{AB}}^2$. Similarly we
have the symmetrized variance $\widehat{\var}(\ket{\psi_{AB}},
\bsA\otimes \bsB)$,
\begin{equation}
\begin{array}{rcl}
\widehat{\var}(\ket{\psi_{AB}}, \bsA\otimes
\bsB)=\frac1{d!}\sum_{\pi\in S_d} \var( \bsP_{\pi}\otimes \bsP_{\pi}
\ket{\psi_{AB}}, \bsA\otimes \bsB).
\end{array}
\end{equation}
By direct calculation we have
\begin{eqnarray*}
&&\widehat{\var}(\ket{\psi_{AB}}, \bsA\otimes \bsB)
= \kappa(\bsA\bsB)\Br{1-\sum^{d-1}_{i=0}\abs{\lambda_i}^4}.
\end{eqnarray*}
Then the standard symmetrized variance
$\widehat{V}(\ket{\psi_{AB}})$ is given by
\begin{eqnarray}
\widehat{V}(\ket{\psi_{AB}})&=& \frac{1}{\kappa(\bsA\bsB)}
\widehat{\var}(\ket{\psi_{AB}}, \bsA\otimes
\bsB)\\&=&1-\sum^{d-1}_{i=0}\abs{\lambda_i}^4\nonumber.
\end{eqnarray}

It is obvious that $\widehat{V}(\ket{\psi_{AB}})$ is a real
symmetric concave function of the Schmidt vector
$(\lambda_0,\lambda_1, \cdots, \lambda_{d-1})^\t$. Therefore,
$\widehat{V}(\ket{\psi_{AB}})$ is an entanglement measure for pure
state $\ket{\psi_{AB}}$ \cite{G. Vidal}. Then the convex roof extension of the
standard symmetrized variance
$\widehat{V}_c(\rho_{AB})=\min_{\set{p_k, \ket{\psi^{(k)}_{AB}}}}
\sum_k p_k \widehat{V}(\ket{\psi^{(k)}_{AB}})$, where the minimum
runs over all possible pure state decompositions of
$\rho_{AB}=\sum_k p_k |\psi^{(k)}_{AB}\rangle \langle
\psi^{(k)}_{AB}|$, is an entanglement measure for bipartite quantum
states. Recall that the well-known entanglement measure concurrence
of a pure state $|\psi_{AB}\rangle$ is defined by
$\mathrm{E}(\ket{\psi_{AB}})=\sqrt{2(1-\Tr{\rho_A^2})}$ with
$\rho_A=Tr_B (|\psi_{AB}\rangle\langle\psi_{AB}|)$, and
$E(\rho_{AB})=\min_{\set{p_k, \ket{\psi^{(k)}_{AB}}}} \sum_k p_k
\mathrm{E}(\ket{\psi_{AB}^{(k)}})$ with the minimum running over all
possible pure state decompositions of $\rho_{AB}=\sum_k p_k
|\psi^{(k)}_{AB}\rangle \langle \psi^{(k)}_{AB}|$ \cite{W.
Wooters,P. Rungta}. By straightforward derivation we find that
$\widehat{V}(\ket{\psi_{AB}})=\frac{1}{2}\mathrm{E}^2(\ket{\psi_{AB}})$
and $\widehat{V}_c(\rho_{AB})=\frac{1}{2}\min_{\set{p_k,
\ket{\psi^{(k)}_{AB}}}} \sum_k p_k
\mathrm{E}^2(\ket{\psi_{AB}^{(k)}})$. Equivalently,
$E(\rho_{AB})=\min_{\set{p_k, \ket{\psi^{(k)}_{AB}}}} \sum_k p_k
\sqrt{2\widehat{V}(\ket{\psi_{AB}^{(k)}})}$, namely, the
entanglement measure concurrence can be expressed in terms of the
variance.

Since the standard symmetrized variance
$\widehat{V}(\ket{\psi_{AB}})$ is observable-independent, we can
choose the projective measurement onto the Schmidt basis of the pure
state $\ket{\psi_{AB}}=\sum_{i=0}^{d-1}\lambda_i \ket{ii}$,
$\Pi^{A}_i=|i\rangle\langle i|$ and
$\Pi^{B}_i=|i\rangle\langle i|$. In this way, we
obtain
\begin{eqnarray*}
&&\widehat{V}(\ket{\psi_{AB}})\\=&&\frac1{d}\sum_{i=0}^{d-1} [ \Innerm{\psi_{AB}}{\Pi^{A}_i \otimes \Pi^{B}_i}{\psi_{AB}} - \Innerm{\psi_{AB}}{\Pi^{A}_i \otimes \Pi^{B}_i}{\psi_{AB}}^2].
\end{eqnarray*}
Therefore, in order to measure the entanglement of a pure state
experimentally, one only needs to measure the expectation values
$\Innerm{\psi_{AB}}{\Pi^{A}_i \otimes \Pi^{B}_i}{\psi_{AB}}$ for
$i=0,1,\cdots, d-1$.

Different from the uncertainty and coherence, entanglement
characterizes the quantum correlations between the subsystems and is
basis-independent. Here we have supposed that the pure state
$\ket{\psi_{AB}}$ is in Schmidt decomposition form and the
observables of the subsystems $A$ and $B$ are diagonal in the
Schmidt basis. Then the standard symmetrized variance
$\widehat{V}(\ket{\psi_{AB}})$ as well as the convex roof extension
$\widehat{V}_c(\rho_{AB})$ are basis independent. They are indeed
entanglement measures.

\section{Conclusions}

In summary, we have proposed the standard symmetrized variance for
pure states and extended it to mixed states in three different ways.
The first extension quantifies the uncertainty in quantum states.
The second convex roof extension quantifies the coherence in quantum
states. The third concave bottom extension quantifies the coherence of
assistance. These quantities have been estimated in detail. Besides,
the standard symmetrized variance also quantifies the entanglement
in bipartite systems and gives an experimental way to evaluate the
entanglement by measurements. In this way, the quantum uncertainty,
coherence and entanglement are closely connected by the standard
symmetrized variance.

\begin{acknowledgments}

This work is supported by the National Natural Science Foundation of
China under grant Nos. 12171044, 11971140, and 12075159,
Open Foundation of State key Laboratory of Networking and Switching Technology (Beijing University of Posts and Telecommunications) (SKLNST-20**-2-0*),
Beijing Natural
Science Foundation (Z190005), Academy for Multidisciplinary Studies,
Capital Normal University, Shenzhen Institute for Quantum Science
and Engineering, Southern University of Science and Technology
(SIQSE202001), and the Academician Innovation Platform of Hainan
Province.

\end{acknowledgments}

\section{Appendices}

{\bf{Appendix A. The proof of Theorem 1.}}

\begin{proof}
First, Let $\delta_{\pi(i),j}$ be the Kronecker delta symbol defined by
$\delta_{\pi(i),j}=1$ if $\pi(i)=j$ and $\delta_{\pi(i),j}=0$ if
$\pi(i)\neq j$. We denote $\bsP_{\pi}=(\delta_{\pi(i),j})$ the
permutation matrix representation of $\pi\in S_d$, the set of all
possible permutations on the set $\set{0,1,\ldots,d-1}$. Then for
any diagonal observable $\bsA$ with diagonal entries $a_i$, we have
the following relations
\begin{eqnarray}\label{eq 1}
\frac1{d!}\sum_{\pi\in S_d}a^2_{\pi(i)} = \frac{\Tr{\bsA^2}}d
\end{eqnarray}
for any $i$, and
\begin{eqnarray}\label{eq 2}
\frac1{d!}\sum_{\pi\in S_d}a_{\pi(i)}a_{\pi(j)} =
\frac{[\Tr{\bsA}]^2-\Tr{\bsA^2}}{d(d-1)}
\end{eqnarray}
for any $i\neq j$, where the summation goes over all possible
permutations.

Now we are ready to prove Theorem~1.
From Eq. \eqref{eq 1}, we have
\begin{eqnarray*}
&&\widehat{\var}(\ket{\psi}, \bsA)\\
&=& \frac1{d!}\sum_{\pi\in
S_d}\Br{\Innerm{\psi}{(\bsP^\dagger_\pi \bsA \bsP_\pi)^2}{\psi} -
\Innerm{\psi}{\bsP^\dagger_\pi \bsA
\bsP_\pi}{\psi}^2}\\
&=&\Innerm{\psi}{\frac1{d!}\sum_{\pi\in S_d}\Pa{\bsA^\pi}^2}{\psi}
-\frac1{d!}\sum_{\pi\in S_d} \Innerm{\psi}{\bsA^\pi}{\psi}^2\\
&=&\frac{\Tr{\bsA^2}}d - \frac1{d!}\sum_{\pi\in S_d}
\Innerm{\psi}{\bsA^\pi}{\psi}^2,
\end{eqnarray*}
where $\bsA^\pi=\bsP^\dagger_\pi \bsA \bsP_\pi$. Note that
$\Innerm{\psi}{\bsA^\pi}{\psi}^2
=\sum^{d-1}_{i,j=0}a_{\pi(i)}a_{\pi(j)}r_ir_j$. We have
\begin{eqnarray*}
&&\frac1{d!}\sum_{\pi\in S_d} \Innerm{\psi}{\bsA^\pi}{\psi}^2
\\=&&\sum^{d-1}_{i=0}\Pa{\frac1{d!}\sum_{\pi\in S_d}a^2_{\pi(i)}}r^2_i
+
\sum^{d-1}_{i\neq j}\Pa{\frac1{d!}\sum_{\pi\in
S_d}a_{\pi(i)}a_{\pi(j)}}r_ir_j.
\end{eqnarray*}
Using Eqs. \eqref{eq 1} and \eqref{eq 2}, we have further
\begin{eqnarray*}
&&\frac1{d!}\sum_{\pi\in S_d} \Innerm{\psi}{\bsA^\pi}{\psi}^2
\\=&&\frac{\Tr{\bsA^2}}d  \sum^{d-1}_{i=0}r^2_i +
\frac{[\Tr{\bsA}]^2-\Tr{\bsA^2}}{d(d-1)}
\Pa{1-\sum^{d-1}_{i=0}r^2_i}.
\end{eqnarray*}
Therefore
\begin{eqnarray*}
\widehat{\var}(\ket{\psi}, \bsA)
&=&
\frac{d\Tr{\bsA^2}-[\Tr{\bsA}]^2}{d(d-1)}\Br{1-\sum^{d-1}_{i=0}r^2_i}.
\end{eqnarray*}
This completes the proof.
\end{proof}

{\bf{Appendix B. The proof of Theorem 2.}}

\begin{proof}
Analogous to the proof of the Theorem~1, one
gets
\begin{eqnarray*}
\widehat{V}(\rho)&=&\frac{1}{\kappa(\bsA)}\frac1{d!}\sum_{\pi\in S_d} \Br{\Tr{\bsA^2\bsP_{\pi} \rho \bsP^\dagger_{\pi}}- \Br{\Tr{\bsA\bsP_{\pi} \rho \bsP^\dagger_{\pi}}}^2}\\
&=&1-\sum^{d-1}_{i=0} \rho_{ii}^2\\
&=&S_L(\Pi(\rho)).
\end{eqnarray*}
This completes the proof.
\end{proof}

{\bf{Appendix C. The proof of the equation
\begin{eqnarray*}
&&\widehat{V}(p_1 \rho_1 \oplus p_2 \rho_2\oplus \cdots \oplus p_n
\rho_n)\\=&& p_1^2\widehat{V}(\rho_1) + p_2^2\widehat{V}(\rho_2)+ \cdots
+ p_n^2\widehat{V}(\rho_n) + \sum_{i\neq j}p_ip_j.
\end{eqnarray*}
}}

\begin{proof}
If $\rho=p_1 \rho_1 \oplus p_2 \rho_2 \oplus \cdots \oplus p_n \rho_n$, by Theorem 2, we have
\begin{eqnarray*}
&&\widehat{V}(p_1 \rho_1 \oplus p_2 \rho_2 \oplus \cdots \oplus p_n \rho_n)\\
&=&1-Tr(\Pi(\rho))^2\\
&=&1-Tr(\Pi(p_1 \rho_1 \oplus p_2 \rho_2 \oplus \cdots \oplus p_n \rho_n))^2\\
&=&1-Tr(\Pi(p_1 \rho_1))^2-Tr(\Pi(p_2 \rho_2))^2- \cdots -Tr(\Pi(p_n \rho_n))^2\\
&=&[p_1^2- p_1^2 Tr(\Pi(\rho_1))^2] + [p_2^2- p_2^2 Tr(\Pi(\rho_2))^2] \\
&&+ \cdots + [p_n^2- p_n^2 Tr(\Pi(\rho_n))^2]+  \sum_{i\neq j}p_ip_j \\
&=& p_1^2\widehat{V}(\rho_1) + p_2^2\widehat{V}(\rho_2)+ \cdots + p_n^2\widehat{V}(\rho_n) +  \sum_{i\neq j}p_ip_j.
\end{eqnarray*}
\end{proof}

{\bf{Appendix D. The proof of Theorem 3.}}

\begin{proof}
For any mixed state $\rho$, by the relation about the
minimum average invariance over all pure state decomposition and the
quantum Fisher information \cite{Yu2013,Toth2013,S. Braunstein,G.
Toth2017}
\begin{equation}\label{eq fisher and var}
F(\rho, \bsA)=4\min_{\set{p_k, \ket{\psi_k}}} \sum_k p_k
\var(\ket{\psi_k}, \bsA) ,
\end{equation}
we have
\begin{eqnarray*}\label{eq var c lower bound}
&&\min_{\set{p_k, \ket{\psi_k}}} \sum_k p_k
\widehat{\var}(\ket{\psi_k}, \bsA)\\\nonumber
&&\geqslant   \frac1{d!}\sum_{\pi\in S_d} \min_{\set{p_k,
\ket{\psi_k}}} \sum_k p_k \var( \bsP_{\pi} \ket{\psi_k},  \bsA
)\notag\\
&&= \frac1{d!}\sum_{\pi\in S_d} \frac{1}{4}
F(\bsP_{\pi} \rho \bsP^\dagger_{\pi}, \bsA)\notag\\
&&= \frac{1}{2} \sum_{k,l}
\frac{(\lambda_k-\lambda_l)^2}{\lambda_k+\lambda_l}
\frac1{d!}\sum_{\pi\in S_d} |\Innerm{ \phi_k }{\bsP^\dagger_{\pi}
\bsA
\bsP_{\pi}}{\phi_l}|^2\notag\\
&&= \frac{1}{2} \sum_{k,l}
\frac{(\lambda_k-\lambda_l)^2}{\lambda_k+\lambda_l}   \sum_{i,j}
\phi^{(k)*}_i \phi^{(l)}_i \phi^{(l)*}_j \phi^{(k)}_j
\Pa{\frac1{d!}\sum_{\pi\in S_d} a_{\pi(i)} a_{\pi(j)}}.\notag
\end{eqnarray*}
For $i=j$, we have by Eq.~\eqref{eq 1},
\begin{eqnarray*}
&&\sum_{i} \abs{\phi^{(k)}_i}^2 \abs{\phi^{(l)}_i}^2
\Br{\frac1{d!}\sum_{\pi\in S_d} a_{\pi(i)}^2} \\=&&
\frac{\Tr{\bsA^2}}{d}\sum_{i} \abs{\phi^{(k)}_i}^2
\abs{\phi^{(l)}_i}^2.
\end{eqnarray*}
For $i\neq j$, from Eq.~\eqref{eq 2} and the orthogonality
$\iinner{\phi_k}{\phi_l}=0$ for $k\neq l$ we have
\begin{eqnarray*}
&&\sum_{i\neq j} \phi^{(k)*}_i \phi^{(l)}_i \phi^{(l)*}_j \phi^{(k)}_j
\Br{\frac1{d!}\sum_{\pi\in S_d} a_{\pi(i)} a_{\pi(j)}}\\
= &&-\frac{[\Tr{\bsA}]^2-\Tr{\bsA^2}}{d(d-1)} \sum_{i}
\abs{\phi^{(k)}_i}^2 \abs{\phi^{(l)}_i}^2.
\end{eqnarray*}
Therefore,
\begin{eqnarray*}
&&\min_{\set{p_k, \ket{\psi_k}}} \sum_k p_k
\widehat{\var}(\ket{\psi_k}, \bsA) \\
\geqslant &&\frac{\kappa(\bsA)}{2}
\sum_{k,l} \frac{(\lambda_k-\lambda_l)^2}{\lambda_k+\lambda_l}
\sum_{i} |\phi^{(k)}_i|^2 |\phi^{(l)}_i|^2,
\end{eqnarray*}
which completes the proof.
\end{proof}

{\bf{Appendix E. The proof of Theorem 4.}}

\begin{proof}
In the permutation group $S_2$, it has two permutations: $(1)$ and $(1,2)$.
For any qubit mixed state $\rho$, one has
$\bsP^\dagger_{\pi}\sigma_3\bsP_{\pi}=-\sigma_3$ for $\pi=(1,2)$. Utilizing
the Eq. (\ref{eq fisher and var}),
we have
\begin{eqnarray*}
&&\min_{\set{p_k, \ket{\psi_k}}} \sum_k p_k
\widehat{\var}(\ket{\psi_k}, \sigma_3)\\
&=&\frac1{2} \min_{\set{p_k, \ket{\psi_k}}} \sum_k p_k \sum_{\pi\in S_2} \var( \bsP_{\pi} \ket{\psi_k},  \sigma_3 )\\
&=& \frac1{2} \min_{\set{p_k, \ket{\psi_k}}} \sum_k p_k \sum_{\pi\in S_2} \var( \ket{\psi_k}, \bsP^\dagger_{\pi}  \sigma_3 \bsP_{\pi} )\\
&=& \min_{\set{p_k, \ket{\psi_k}}} \sum_k p_k \var( \ket{\psi_k},\sigma_3 )\\
&=&\frac{1}{4} F(\rho, \sigma_3)\\
&=&(\lambda_1-\lambda_2)^2 \Innerm{\phi_1}{\sigma_3}{\phi_2}^2.
\end{eqnarray*}
Combining with $\kappa(\sigma_3)=2$, we get
$\widehat{V}_c(\rho)=\frac{1}{8} F(\rho,
\sigma_3)=\frac{1}{2}(\lambda_1-\lambda_2)^2
\Innerm{\phi_1}{\sigma_3}{\phi_2}^2$.
This completes the proof.
\end{proof}

{\bf{Appendix F. The proof of Theorem 5.}}

\begin{proof}
In qubit systems, we have $\widehat{V}_c(\rho)=\frac{1}{8} F(\rho,
\sigma_3)$ and $\widehat{V}_a(\rho)=\widehat{V}(\rho)=\frac{1}{2}
\var(\rho, \sigma_3)$. For any rank two mixed state $\rho$ and
observable $\bsA$, one has $\var(\rho,\bsA)-F(\rho,
\bsA)=\frac{1}{2}[1-\Tr{\rho^2}](\omega_1-\omega_2)^2$ with
$\omega_k$ the eigenvalues of $\bsA$ \cite{G. Toth2017}. Therefore,
we have
\begin{eqnarray*}
\widehat{V}_a(\rho) -\widehat{V}_c(\rho)
&=& \frac{3}{8}\var(\rho,\sigma_3)+\frac{1}{8}[\var(\rho,\sigma_3)-F(\rho, \sigma_3)]\\
&=& \frac{3}{8}[1-\Tr{\Pi(\rho)^2}] +\frac{1}{4}[1-\Tr{\rho^2}].
\end{eqnarray*}
This completes the proof.
\end{proof}


\end{document}